\documentclass[twocolumn,showpacs,preprintnumbers,amsmath,amssymb]{revtex4}

\usepackage{graphicx}% Include figure files
\usepackage{dcolumn}% Align table columns on decimal point
\usepackage{bm}% bold math

\input  epsf 
\def\bea{\begin{eqnarray}}
\def\eea{\end{eqnarray}}
\def\be{\begin{equation}}
\def\ee{\end{equation}}
\def\lan{\langle}
\def\ran{\rangle}

\def\mysize{5.9cm}

\begin{document}

\title{Glass transition as a decoupling-coupling mechanism of rotations}

\author{A.\ D\'\i az-S\'anchez }  
\altaffiliation[Also at ]{Dipartimento di Scienze Fisiche, Universit\`a di Napoli
``Federico II'',  Complesso Universitario di Monte Sant'Angelo,
Via Cintia , I-80126 Napoli, Italy and
INFM, Unit\`a di Napoli, Napoli, Italy.}
\affiliation{
Departamento de F\'\i sica Aplicada,
Universidad Polit\'ecnica de Cartagena, \\ 
Campus Muralla del Mar, Cartagena, 
E-30202 Murcia, Spain.}
\email{andiaz@upct.es}

\date{\today}

\begin{abstract}
We introduce a three-dimensional lattice gas model to study the glass 
transition. In this model the interactions come from the excluded volume
and particles have five arms with an asymmetrical shape,
which results in geometric frustration that inhibits full packing.
Each particle has two degrees of freedom, the position and the orientation 
of the particle. We find a second order phase transition at a density 
$\rho\approx 0.305$, 
this transition decouples the orientation of the particles which can rotate
without interaction in this degree of freedom until $\rho=0.5$ is
reached. Both the inverse diffusivity and the relaxation time follow a power 
law behavior for densities $\rho\le 0.5$. The crystallization at $\rho=0.5$
is avoided because frustration lets to the system to reach higher
densities, then the divergencies are overcome.
For $\rho > 0.5 $ the orientations of the particles are coupled and
the dynamics is governed by both degrees of freedom. 
\end{abstract}
\maketitle

% INTRODUCTION
 
\section{Introduction}

In the last years, a great deal of work has been done to obtain a fundamental
understanding of the glass transition. Many questions about the 
equilibrium and the dynamical properties of the glassy state remain ananswered.
It is not clear if there is a true phase transition and what is the role that 
geometric frustration plays on it. The relations between the equilibrium 
and the dynamical properties are not understood \cite{M01}. 
The mode coupling theory for supercooled liquids \cite{G91} predicts the 
existence of a temperature $T_{\rm c}$ at which there is a crossover from a 
liquid to a glassy state. In the glassy state the dynamics would be dominated 
by complex activated processes.
For temperatures $T>T_{\rm c}$ but close to $T_{\rm c}$ there is a power law 
behavior of the relaxation time and also of the inverse diffusivity. 
It is not understood whether $T_{\rm c}$ is a purely kinetic transition 
temperature \cite{G91} or if it is a true thermodynamic glass transition 
which is kinetically avoided \cite{DS01}.
Several lattice gas models have been used to simulate glassy systems
and have reproduced some aspects of the glassy phenomenology.
For example the Hard Square Model (HSM) \cite{GF66}, the Kinetically Constrained 
Model \cite{FA84,GP97}, the Frustrated Ising Lattice Gas Model \cite{NC98}, the
one introduced by Ciamarra {\sl et al.} \cite{C02},
and recently the Lattice Glass Model (LGM) \cite{BM01}. The LGM model with density 
constraint $l=0$ is equivalent to the the three-dimensional HSM model also called 
Hard Cubic Model (HCM). The LGM model relates the glass transition to a first order 
phase transition.

In this paper we consider a three-dimensional lattice gas model,
which contains as main ingredients only geometric frustration without
quenched disorder and without kinetic or density constraints, as quenched disorder
is not appropriate to study structural glasses and kinetic or density constraints 
are some how artificial. Similar models have already been proposed and studied
in two-dimensional systems \cite{C94,BK01,DD02} and applied to study granular 
material \cite{CL97}.

\begin{figure}
\epsfysize=\mysize
\begin{center}
\epsfbox{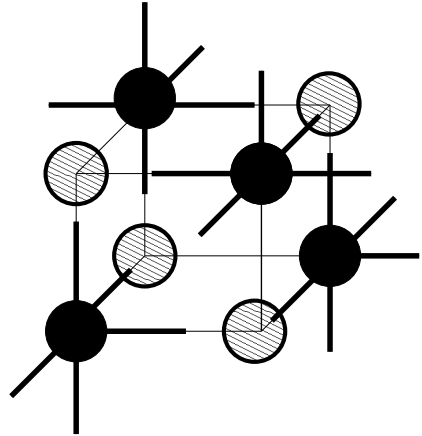}
\end{center}
\caption{ Schematic picture of one particular configuration in
a system size of $N=2^3$ and density $\rho=0.5$. Shadow spheres represent
holes in the system. Black spheres are particles with five arms each one.
} 
\label{fig:1}
\end{figure} 

% MODEL

\section{The model}

Our model is a generalization to three dimensions of the two-dimensional 
model studied in Ref. \cite{DD02}. It can be considered 
as an illustration of the concept of frustration arising as a packing problem. 
We have particles with five arms and they occupy the vertices of
a cubic lattice with one of six possible orientations. Assuming that the 
arms cannot overlap due to excluded volume, we see that only for some 
relative orientations two particles can occupy nearest-neighbor vertices.
Consequently, depending on the local arrangement of particles,
there are sites on the lattice that cannot be occupied (see Fig. \ref{fig:1}).
This type of ``packing'' frustration thus induces defects or 
holes in the system. We impose periodic boundary conditions in the 
cubic lattice of size $N=L^3$. The maximum of density is $\rho_{\rm max}=3/5$ 
at which all possible bonds are occupied by an arm. Here we have
two degrees of freedom for each particle, the position and the orientation 
of the particle. This model is the HCM model when the particles have six instead 
five arms. Our model would be also similar to the LGM model with the density 
constraint $l=1$. We will compare the results found in our model with the ones 
obtained in these two models.
We have used two algorithms in order to make the simulations.
The first one (CA) is the Monte Carlo simulation at fixed density in the canonical
ensemble, we have simulated the diffusion and rotation dynamics of the particles 
by the following algorithm:
i) Pick up a particle at random;
ii) Pick up a site at random between the six nearest neighbor ones;
iii) Choose randomly an orientation of the particle;
iv) If it does not cause the overlapping of two arms,
move the particle in the given site with the given orientation;
v) If the diffusion movement is not possible,
choose a random orientation and try to rotate the particle to this new
orientation;
vii) Advance the clock by $1/N$, where $N$ is the number
of sites, and go to i). The second algorithm (GCA) is the grand canonical 
ensemble, the diffusion and rotation dynamics is as in the CA simulations
but now a reservoir with chemical potential
$\mu$ is coupled to each lattice site which can create (if it does not cause the 
overlapping of two arms) or destroy particles. As we expect
the GCA simulation reaches the equilibrium faster. We will use the GCA simulation
in order to find the behavior of the density $\rho$ with $\mu$ \cite{BM01}.

% DIFFUSION
\section{Results}

\begin{figure}
\epsfysize=\mysize
\begin{center}
\epsfbox{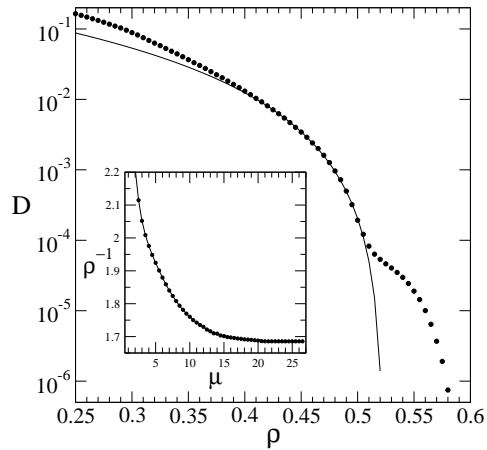}
\end{center}
\caption{Diffusion constant $D$ as a function of density $\rho$ for 
a system size of $N=14^3$. The fitting function is a power law 
$D=2.11(0.523-\rho)^{2.45}$. Inset: The inverse of the density as 
function of $\mu$ on a lattice of size $N=14^3$. The data are obtained
from the GCA with 2$\times 10^7$ Monte Carlo Steps per Particle at a fixed
increasing rate of the chemical potential between $\mu = 0$ and $\mu=27$.
} 
\label{fig:2}
\end{figure}

We first study a possible first order phase transition in our model.
In the inset of Fig. \ref{fig:2} the inverse of the density is plotted as a
function of the chemical potential. As in Ref. \cite{BM01} we make GCA 
simulations to obtain this figure. A maximum of density very close to 
$\rho_{\rm max}$ is reached without any discontinuity, although 
we observe finite-size effects which prevent to reach $\rho_{\rm max}$ for the 
lattice sizes studied here. So, first order phase transition is not present in 
our system. In the HCM model we also observe similar behavior of the density with 
$\mu$ reaching the maximum of density continously, $\rho_{\rm max}=0.5$ 
in the HCM model. Instead, in the LGM model with $l \ge 1$ there is always a 
first order phase transition \cite{BM01}.

We now calculate the diffusion coefficient $D$ from the mean-square 
displacement of the particles at very long times with the CA simulations. 
The values obtained for $D$ are well fitted by a power law close to 
$\rho_{\rm c}=0.52 \pm 0.005$ and for densities lower than $\rho=0.5$, 
$D\propto (\rho_{\rm c}-\rho)^\gamma$ with $\gamma=2.45\pm 0.01$ 
(see Fig. \ref{fig:2}). 
This anomalous behavior of $D$ near of $\rho_{\rm c}$ would indicate
a crossover density, from liquid to glass phase, where activated 
processes dominate in the glass phase. 
As it happens in the HSM model 
\cite{F89} the finite size effects for the diffusivity in
the HCM model are very large when we do CA simulations, it is because the particles 
can be enclosed in cages and the diffusion is blocked. It prevents to reach 
densities close to the maximum density. Nevertheless, in our model the 
finite size effects are only important for $\rho > 0.58$ when $N=14^3$. This is 
because our model has two degrees of freedom and rotations prevent to find blocked 
configurations for $\rho < 0.58$.  

\vglue 0.5cm

% ORDERED PHASE

\begin{figure}
\epsfysize=\mysize
\begin{center}
\epsfbox{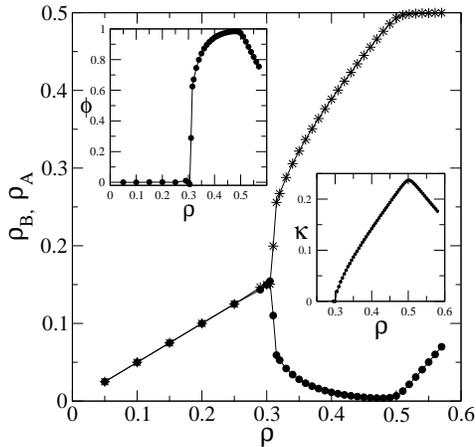}
\end{center}
\caption{Concentration of particles in sublattices $A$, $\rho_A$ (asterisks),
and $B$, $\rho_B$ (solid circles), as a function of the density 
$\rho$ for a system size of $L=14$. Left inset: Order parameter $\phi$ as a 
function of the density $\rho$. Right inset: Compressibility $\kappa$ as a
function of the density $\rho$.
}
\label{fig:3}
\end{figure}

In order to understand what happens at $\rho \approx 0.5$ 
we study the following  microscopic order parameter.
As in anti-ferromagnetic systems, the cubic lattice is divided
into two interpenetrating sublattices ($A$ and $B$), 
a site in a sublattice has six nearest neighbor sites which belong to
the other sublattice. The order parameter is defined as
\be
\phi=\frac{\rho_A-\rho_B}{\rho_A+\rho_B} \, ,
\ee
where $\rho_A$ and $\rho_B$ are the equilibrium concentrations of the particles in
the sublattices $A$ and $B$ and we have $\rho=\rho_A+\rho_B$.
This parameter can be used to study concentration sublattice ordering. 
When the particles prefer to stay in one of these sublattices then $\phi \ne 0$.
In the left inset of Fig. \ref{fig:3} we show $\phi$ as a function of the density.
We can see that it is different to zero for $\rho >\rho_{\rm f}\approx 0.3$ 
and it increases until $\rho \approx 0.5$, then there is a maximum. 
For higher densities it decreases linearly with the density. 
In Fig. \ref{fig:3} we show the concentration in both sublattices. We see that
at $\rho_{\rm f}$ the concentrations $\rho_A$ and $\rho_B$ begin to be different
each other, the particles prefer to stay in a sublattice. The concentration in the 
sublattice $A$ has the maximum value when $\rho=0.5$ is reached. A sublattice is 
full of particles at this concentration while the other one is in practice empty.
The particles begin to occupy the empty sublattice for densities higher than 
$\rho=0.5$ remaining the other sublattice full, then the parameter $\phi$ 
decreases. Here we observe that there are not frozen particles at these densities,
$\rho_A$ and $\rho_B$ are equilibrium concentrations. The order parameter
$\phi$ can take positive and negative values, it depends on which sublattice has
higher density for $\rho >\rho_{\rm f}$, in Fig. \ref{fig:3} we have $\phi > 0$
for $\rho>\rho_{\rm f}$ because $\rho_A > \rho_B$. 
Similar behavior for the parameter $\phi$ is found in the HCM model but 
$\rho_{\rm f}\approx 0.22$ is lower than the one obtained in our model 
and the system crystallizes at $\rho=0.5$, its maximum of density, then the 
particles are frozen in a sublattice and it is not possible to reach higher 
densities.  So, the last part of Fig. \ref{fig:3} (for $\rho>0.5$) is not found 
in the HCM model. As we will see below at $\rho_{\rm f}$ there is a continuous phase 
transition in our model and also in the HCM model. In the LGM model with $l= 1$, 
we have found a first order phase transition with a discontinuity of the density,
when is plotted as a function of $\mu$, from a density $\rho \approx 0.32$ to 
a density $\rho = 0.5$, then the system crystallizes and the particles are frozen in a 
sublattice.

\vglue 0.5cm

% COMPRESSIBILITY

\vglue 0.5cm

\begin{figure}
\epsfysize=\mysize
\begin{center}
\epsfbox{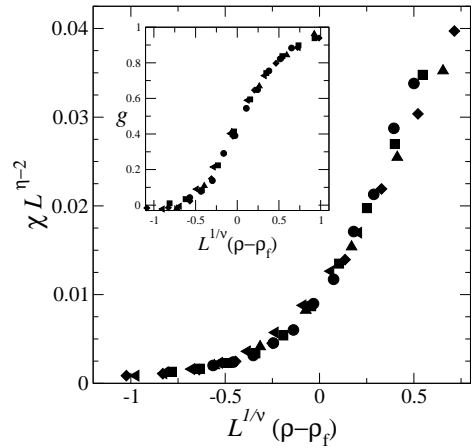}
\end{center}
\caption{Finite size scaling of $\chi$ and $g$ (inset) for lattice
sizes $L=8$ ($\bullet$), 10 ($\blacksquare$), 12 ($\blacklozenge$),
14 ($\blacktriangle$), and 16 ($\blacktriangleleft$). 
$\rho_{\rm f} = 0.305\pm0.005$, $\eta = 0.04 \pm0.01$, and 
$\nu= 0.63 \pm 0.05$. 
}
\label{fig:4}
\end{figure}

We now study the equilibrium second order phase transition in our system.
For that, we define the compressibility by the following expression
\be
\kappa=\frac{1}{N} \sum_i \left(\lan n_i^2\ran-\lan n_i\ran^2 \right) \,,
\ee
where $n_i=0,1$ is the occupation number of site $i$ and $\lan\cdots\ran$ 
indicates equilibrium average.
In the right inset of Fig. \ref{fig:3} we see the behavior of  $\kappa$ with the 
density. We find that the compressibility is different to zero for densities 
higher than $\rho_{\rm f} \approx 0.305$ and it has a maximum at $\rho = 0.5$, 
decreasing for higher densities. This behavior is similar to the one found for 
the parameter $\phi$. The associated susceptibility $\chi$ is given by
$ \chi= N \left(\langle \kappa^2\rangle -\langle \kappa\rangle ^2\right)$
and the Binder's cumulant is $g=(3-\langle \kappa^4\rangle/\langle 
\kappa^2\rangle^2)/2$.
Around a continuous phase transition $\chi$ and $g$ should obey the 
finite-size scaling
$\chi(\rho)= L^{2-\eta}\widetilde{\chi}\left[L^{1/\nu}(\rho-\rho_{\rm f})\right]$
and  $g(\rho)=\widetilde{g}\left[L^{1/\nu}(\rho-\rho_{\rm f})\right]$
where $\widetilde{\chi}\left[x\right]$ and $\widetilde{g}\left[x\right]$
are universal functions and $\rho_{\rm f}$ is the critical density.
From the finite size scaling (see Fig. \ref{fig:4}) 
we find a continuous phase transition at 
$\rho_{\rm f} = 0.305\pm0.005$ which belongs to the three-dimensional 
Ising universality class, $\eta = 0.04 \pm0.01$ and $\nu= 0.63 \pm 0.05$. 
In the HCM model we have found a second order phase transition 
which also belong to the same universality class but with 
$\rho_{\rm f} \approx 0.22$. 
In Fig. \ref{fig:2} we can see that the diffusion constant
is not affected by the second order phase transition.

% CORRELATIONS

\vglue 0.5cm

\begin{figure}
\epsfysize=\mysize
\begin{center}
\epsfbox{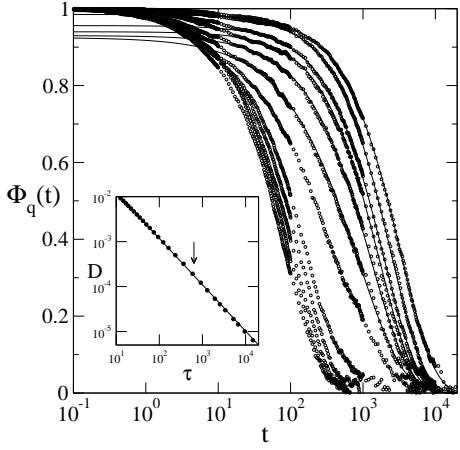}
\end{center}
\caption{Correlation functions of the density fluctuations $\Phi_{\bf q}(t)$ for 
${\bf q}=(\pi/3,\pi/3,\pi/3)$, system size $L=14$, and densities (from bottom to top)
$\rho = 0.45$, 0.46, 0.47, 0.48, 0.49, 0.5, 0.505, 0.51, 0.515, 0.52, 0.53, and
0.54. The solid lines are fitting functions corresponding to stretched
exponential functions. For $\rho \ge 0.515$ we have $\beta=0.94 \pm0.01$. 
Inset: $D$ as a function of the relaxation time $\tau$. The arrow shows the
density $\rho =0.5$. The solid line is the fitting function $D=0.135/\tau$.
}
\label{fig:5}
\end{figure}

We now study the relaxation of the autocorrelation functions of the density
fluctuations  
\be
\Phi_{\bf q}(t)=\frac{\lan \rho_{\bf q}^*(t'+t) \rho_{{\bf q}}(t')\ran}
{\lan|\rho_{\bf q}|^2\ran}\,, 
\label{corre}
\ee
where $\lan\cdots\ran$ denotes average over the reference time $t'$ and
$\rho_{\bf q}$ is the Fourier transform on the lattice of the density
\be
\rho_{\bf q}(t)=\frac{1}{N}\sum_{i=1}^n e^{-i{\bf q}.{\bf r}_i(t)} \,,
\ee
where ${\bf r}_i(t)$ is the position of the $i$th particle at time $t$,
$n$ is the number of particles and ${\bf q}$ is the wave number. Because of
the periodic conditions on the cubic lattice ${\bf q}=(2\pi/L)(n_x,n_y,n_z)$,
with $n_x,n_y,n_z=1 ,\ \cdots ,\  L/2$.
Fig. \ref{fig:5} shows $\Phi_{\bf q}(t)$ for ${\bf q}=(\pi/3,\pi/3,\pi/3)$ and 
different densities. We can see a two-step relaxation decay for 
$\rho > 0.5$, the second relaxation step can be fitted by a stretched exponential 
form, $f(t)=f_0\exp\left[(t/\tau)^\beta\right]$ where the exponent 
$\beta=0.94 \pm0.01$ remains constant with $\rho$ for $\rho \ge 0.515$. 
For $\rho < 0.49$ we only have a one-step relaxation decay.
The relaxation time $\tau$ can be obtained from $\Phi_{\bf q}(t)$.
We find that it is proportional to the inverse of the diffusivity, 
$\tau \propto D^{-1}$, for the whole range of densities studied here (see inset of
Fig \ref{fig:5}). Thus, the relaxation time follows a power law at $\rho <\rho_{\rm c}$ 
with the same exponent than the one obtained in the power law of $D^{-1}$ (see 
Fig. \ref{fig:2}).

\begin{figure}
\epsfysize=\mysize
\begin{center}
\epsfbox{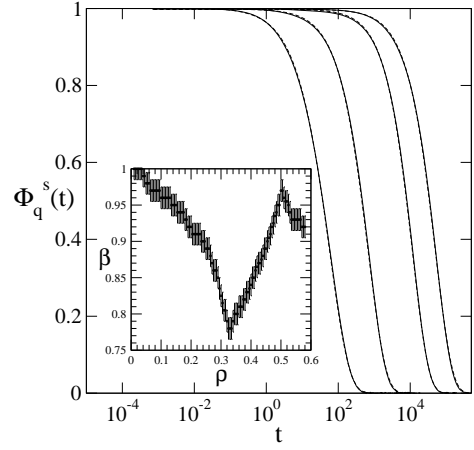}
\end{center}
\caption{Self-part $\Phi_{\bf q}^s(t)$ of the density-density
autocorrelations for ${\bf q}=(\pi,0,0)$ and densities 
$\rho=0.35$, 0.45, 0.5, and 0.53.
Dotted lines are fitting functions corresponding to
stretched exponential functions.
Inset: Parameter $\beta$ of the stretched exponential
function for $\Phi_{\bf q}^s(t)$ as a function of
the density (for each density we plot the error bar of $\beta$).
  }
\label{fig:6}
\end{figure}

We now present the results for the self-part of the autocorrelation function 
of the density fluctuations, defined as
\be
\Phi_{\bf q}^s(t) =\frac{1}{N \rho}\sum_i
\lan e^{i{\bf q }({\bf r}_i(t'+t)-{\bf r}_i(t'))}\ran,
\ee
where $\lan\cdots\ran$ denotes average over the reference time $t'$.
Fig. \ref{fig:6} shows $\Phi_{\bf q}^s(t)$ corresponding to 
${\bf q}=(\pi,0,0)$ for densities $\rho= 0.35$, 0.45, 0.5, 
and 0.53. For the whole range of densities studied here we find that the 
whole time interval of $\Phi_{\bf q}^s(t)$ can be fitted by a stretched 
exponential function where the exponent $\beta$ depends on the
density. In the inset of Fig. \ref{fig:6} we show $\beta$ as a
function of the density. The exponent $\beta$ decreases with the 
density until a density near $\rho_{\rm f}$ is reached. Starting from 
this density, which corresponds to the second order phase transition, 
the exponent increases until $\rho=0.5$ is reached. For 
$\rho > \rho_{\rm c}$ it becomes constant (within the error bars)
$\beta \approx 0.93$. The relaxation time obtained from the fit of
$\Phi_{\bf q}^s(t)$ is proportional to the inverse of the diffusivity.

\vglue 0.5cm

\begin{figure}
\epsfysize=\mysize
\begin{center}
\epsfbox{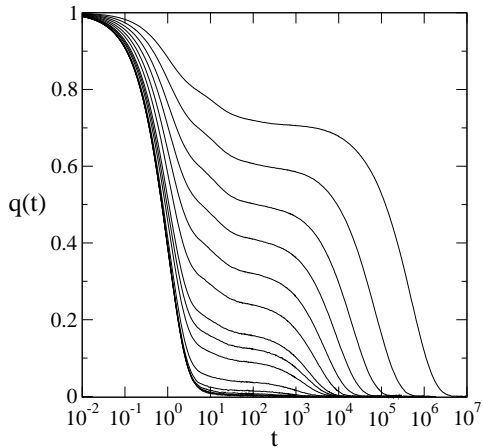}
\end{center}
\caption{Relaxation functions of the self-overlap $q(t)$ for system size 
$L=14$ and densities (from bottom to top) $\rho = 0.45$, 0.46, 0.47, 0.48, 
0.49, 0.5, 0.51, 0.515, 0.52, 0.53, 0.54, 0.55, 0.56, 0.57, and 0.58.
}
\label{fig:7}
\end{figure}

In order to study the role of the orientation of the particles we define a 
self-overlap parameter similar to that defined in \cite{DF99} for liquids
but which also takes into account the orientation of the particles, besides
their position. The orientation of a particle is defined by the discrete values
of the two orthogonal angles $\theta_i = 0,\ \pi/2,\ \pi$, or $3\pi/2$ and 
$\varphi_i = 0,\ \pi/2,\ \pi$, or $3\pi/2$. We define the self-overlap as
\[
q(t)=\frac{1}{N\rho}\sum_{i}^n \lan n_i(t')n_i(t'+t)
\]
\be
\cos\left[\theta_i(t'+t)-\theta_i(t')\right] 
\cos\left[\varphi_i(t'+t)-\varphi_i(t')\right] \ran \,,
\ee
here $\lan\cdots\ran$ denotes average over the reference time $t'$.  
If all the particles have the position and the orientation frozen 
then $q(t)=1$. Fig. \ref{fig:7} shows the parameter $q(t)$ for different 
values of the density. The plateau becomes 
visible for densities higher than $\rho \approx 0.5$. From this density
there are an important number of particles which have frozen 
the position and the orientation for a long time. The number of frozen particles and
the time during they are frozen increase with the density. We can fit the second 
relaxation step with the stretched exponential function, but now the exponent
$\beta$ decreases with the density from $\beta = 0.94$ for $\rho = 0.52$ until
$\beta = 0.77$ for $\rho = 0.58$.

\vglue 0.5cm

% CONCLUSION

\section{Conclusion}

We have proposed a three-dimensional lattice gas model, based on the concept of
geometric frustration which is generated by the particle shape. In this model
a second order phase transition decouples the orientation of the particles
which can rotate without interaction in the orientation degree of 
freedom until $\rho=0.5$ is reached. This is because in practice 
the particles remain all the time in a sublattice and then the particles can rotate 
freely. For densities higher than $\rho=0.5$ geometric frustration 
begins to work and rotations are governed by complex collective processes. 
Then, the two degrees of freedom are important in the diffusivity movement of 
the particles. For $\rho \le 0.5$ the system is going to a crystalline state 
with all the particles frozen in a sublattice, this brings to a power law divergency 
of the relaxation time and the inverse of diffusivity for $\rho < 0.5$.
But frustration lets to the system reach higher densities and crystallization
is avoided and the divergencies are overcome. Then, vibrational effects  
are observed which bring to the two-step relaxation decay in the density 
correlations and in the self-overlap parameter. Thus, the glass transition is purely 
a kinetic transition in our model. Geometric frustration plays a fundamental role,
without frustration the arrest would be close to $\rho_{\rm c}$, but also 
the second order phase transition is very important, which decouples the 
orientation of the particles. In the two-dimensional model \cite{DD02}, 
which has not second order phase transition, we do not observe these anomalies 
in the diffusivity and relaxation time. The order parameters $\phi$
and $\kappa$ exhibit a maximum at the glass transition $\rho\approx 0.5$.  

We have found that the diffusion constant is not affected by the second
order phase transition. The self-part of the autocorrelation function
of the density fluctuation can be fitted by a stretched exponential
function with an exponent $\beta$ that has a minimum value at the
second order phase transition and a local maximum at the glass transition.

% ACKNOWLEDGMENTS

\begin{acknowledgments}
This work was supported in part by the European TMR Network-Fractals
(Contract No. FMRXCT\-980183) and Project No. Pi-60/00858/FS/01 from
the Fundaci\'on S\'eneca Regi\'on de Murcia. 
A part of it was performed during a postdoctoral visit at the 
Universit\`a di Napoli ``Federico II''; I thank the Universit\`a 
di Napoli ``Federico II' for its hospitality and the European TMR 
Network-Fractals for a postdoctoral grant. I am indebted to A. Coniglio
for suggesting this type of models.
\end{acknowledgments}

\pagebreak

\end{document}